\documentclass[apj]{emulateapj}
\usepackage{mathptmx}

\def\gtorder{\mathrel{\raise.3ex\hbox{$>$}\mkern-14mu
             \lower0.6ex\hbox{$\sim$}}}
\def\ltorder{\mathrel{\raise.3ex\hbox{$<$}\mkern-14mu
             \lower0.6ex\hbox{$\sim$}}}

\slugcomment{Accepted to ApJ}

\shorttitle{KBOs observations in SDSS}

\shortauthors{Ofek}

\begin{document}

\title{SDSS Observations of Kuiper Belt Objects: Colors and Variability}
\author{Eran~O.~Ofek\altaffilmark{1}$^{,}$\altaffilmark{2}$^{,}$\altaffilmark{3}}

\altaffiltext{1}{Division of Physics, Mathematics and Astronomy, California Institute of Technology, Pasadena, CA 91125, USA}
\altaffiltext{2}{Einstein Fellow}
\altaffiltext{3}{Benoziyo Center for astrophysics, Weizmann Institute of Science, 76100 Rehovot, Israel.}

\begin{abstract}

Colors of Trans Neptunian Objects (TNOs) are used to study the evolutionary
processes of bodies in the outskirts of the
Solar System, and to test theories regarding their origin.
Here I describe a search for serendipitous Sloan Digital Sky Survey (SDSS)
observations of known TNOs and Centaurs.
I present a catalog of SDSS photometry, colors and astrometry of 
388 measurements of 42 outer Solar-System objects.
I find a weak evidence, at the $\approx2\sigma$ level (per trial),
for a correlation
between the $g-r$ color and inclination of
scattered disk objects and hot classical KBOs.
I find a correlation
between the $g-r$ color and the angular momentum in the $z$ direction
of all the objects in this sample.
These findings should be verified using
larger samples of TNOs.
Light curves as a function of phase angle are constructed for 13 objects.
The steepness of the slopes of these light curves
suggests that the coherent backscatter mechanism plays a major role
in the reflectivity of outer Solar-System small objects
at small phase angles.
I find a weak evidence for an anti-correlation,
significant at the $2\sigma$ confidence level (per trial),
between the
$g$-band phase angle slope parameter
and the semi-major axis, as well as the aphelion distance, of these objects
(i.e., they show a more prominent ``opposition effect'' at smaller
distances from the Sun).
However, this plausible correlation should be verified using larger sample.
I discuss the origin of this possible correlation and argue that if this
correlation is real it probably indicates that ``Sedna''-like objects
have a different origin than other classes of TNOs.
Finally, I identify several objects with large variability amplitudes.

\end{abstract}

\keywords{
solar system: general ---
solar system: Kuiper Belt ---
techniques: photometric}

\section{Introduction}
\label{sec:Intro}

Colors and variability of small bodies in the outer Solar System
provide insight into the physical properties and evolution
of these objects.
The colors of these objects are believed to be related
to evolutionary processes such as collisions, resurfacing
and the interaction of cosmic rays with the surface of these
bodies (e.g., Cooper et al. 2003; see however Porter et al. 2010).
The large body of color observations of Trans Neptunian Object (TNOs;
e.g., Luu \& Jewitt 1996;
Delsanti et al. 2001;
Hainaut \& Delsanti 2002;
Trujillo \& Brown 2002;
Tegler \& Romanishin 2003;
Almeida et al. 2009;
Santos-Sanz et al. 2009;
Romanishin et al. 2010;
Sheppard 2010)
is not entirely consistent with theoretical
ideas (see the review in Jewitt, Morbidelli \& Rauer 2008).
The main characteristic of TNO colors is diversity.
To date, the only secure correlation
involving TNO colors 
is between the $B-I_{{\rm c}}$ color and the orbital inclination
of classical Kuiper Belt Objects (KBOs)
and scattered disk objects\footnote{Here TNOs are defined as objects
with semi-major axis larger than that of Neptune. KBOs and scattered disk
objects are loosely defined and here we follow the definition of
Morbidelli \& Brown (2004).}
(e.g., Hainaut \& Delsanti 2002; Trujillo \& Brown 2002; Peixinho et al. 2008).
I note that the reported correlations between
the inclination and $V-R_{{\rm c}}$ or $R_{{\rm c}}-I_{{\rm c}}$ colors
remain controversial (e.g., Stephens et al. 2003).

KBO variability is related to shape, binarity
and albedo surface uniformity.
Measuring the binary frequency allows of testing models of
KBO binary formation (e.g. Goldreich, Lithwick \& Sari 2002;
Weidenschilling 2002).
Moreover, in some cases binaries are used to determine
masses (e.g., Noll et al. 2004) and densities
(e.g., Sheppard \& Jewitt 2004; Gnat \& Sari 2010) of KBOs.

TNO variability studies
typically require medium size telescopes and are therefore
observationally demanding.
The Sloan Digital Sky Survey (SDSS; York et al. 2000)
provides imaging in the $ugriz$-bands of a considerable fraction of the
celestial sphere.
The photometric calibration of the SDSS is good to $\approx1$\%
in the $griz$ bands and $\approx2$\% in the $u$ band
(e.g., Tucker et al. 2006; Padmanabhan et al. 2008).
The SDSS astrometric accuracy is $\approx0.1''$ (e.g., Pier et al. 2003).
However, given the short time interval within which
the SDSS images were obtained
($\approx5$\,min), it does not allow in most cases the detection
of KBO motion\footnote{At opposition, the typical geocentric on sky motion of a Solar-System object orbiting the Sun at 40\,AU is $\sim 0.25''$ in 5\,min}.
Ivezic et al. (2001) and Juric et al. (2001) constructed a catalog of all the SDSS
sources displaying a significant motion within the 5\,min exposures---the
SDSS Moving Object
Catalog\footnote{http://www.astro.washington.edu/users/ivezic/sdssmoc/sdssmoc.html} (SDSSMOC).
However, in the
fourth release of this catalog (SDSSMOC4)
there are only 33 entries of known objects with $a>10$ AU.

Here I describe a search for known small objects in the outer parts of the Solar System
in the existing SDSS imaging data. A compilation of the photometric and astrometric
properties of these bodies are presented and analyzed.
The structure of this paper is as follows.
In \S\ref{DB} I describe the catalog of SDSS observations of
outer Solar-System objects.
In \S\ref{Color} I discuss their colors, while in \S\ref{Var}
I describe their variability properties.
Finally, I summarize the results in \S\ref{Sum}.

\section{A catalog of SDSS observations of TNOs}
\label{DB}

This section describes the construction
of a catalog of SDSS observations of known outer Solar-System objects
with semi-major axes $a>10$\,AU.

\subsection{SDSS images whose footprints may contain known TNOs}
\label{Im}

I retrieved 
a list of the orbital elements of all known (numbered and unnumbered) minor planets
in the Solar System\footnote{http://ssd.jpl.nasa.gov/?sb\_lem}
from the
Jet Propulsion Laboratory (JPL) Horizons\footnote{http://ssd.jpl.nasa.gov/?horizons} system
(updated for 2010 August 2).
Then, I selected
all the objects with semi-major axis $a>10$\,AU.
This list contains 1469 bodies.

I used the SDSS (York et al. 2000) CasJobs\footnote{http://casjobs.sdss.org/casjobs/}
utility to generate a catalog of all the
images\footnote{1,589,702 images.} which are available in the SDSS database.
Here, an ``image'' is defined uniquely by
the SDSS run, rerun, camcol (camera column), and
field\footnote{see definitions in: http://www.sdss.org/dr7/glossary/index.html}.
The catalog contains all the images included in
the SDSS data release 7 (DR7), Segue,
and Stripe 82 (Abazajian et al. 2009).
For each image I obtained the time at which it was observed and
I calculated the coordinates of its four
corners\footnote{Performed by transforming the SDSS great circle coordinates to equatorial coordinates; http://www.sdss.org/dr7/products/general/astrometry.html}.

Next, I used the JPL Horizons system to generate daily ephemerides for each
of the 1469 objects between Julian Day (JD) 2451070 and 2454467.
This JD range contains all the SDSS observations in DR7.
For each entry in the daily ephemerides
of each object I checked whether it is contained within any
of the polygons defining the corners of all the SDSS images
taken within one day of the ephemeris entry.
If a match was found, then the object ephemeris was re-generated
for the exact time at which the image was taken (to an accuracy of 1\,min).
In total, 4642 possible observations of 574 outer Solar-System objects were found.
Of these 845 entries are of objects
with a predicted $V$-band magnitude,
at the time of observation,
brighter than $V_{{\rm pred}}=22$\,mag.

\subsection{Photometry and astrometry of TNOs in SDSS images}
\label{PhotoAst}

Next, I searched for sources in the SDSS
images near the predicted position of the outer Solar-System objects.
Unlike ``typical'' minor planet surveys, this method only yields
a single image per object per field, so one cannot use the motion
of the object between two images of the same field to verify whether it is indeed a
Solar-System object
(rather then a variable star or a transient).
Therefore, as described below,
I exercised great care to remove
false identifications
or contamination by nearby sources.

\begin{figure}
\centerline{\includegraphics[width=8.5cm]{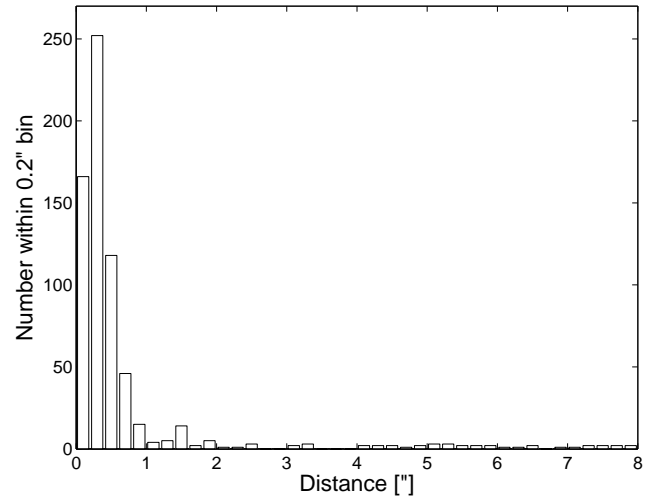}}
\caption{Histogram of angular distances
between the predicted position of the Solar-System objects and the
position of the SDSS source nearest to the predicted position
(either the Solar-System object or a background object).
This plot is shown for 672 measurements in
which I find at least one SDSS source within $8''$ of the predicted
position of the object.
\label{fig:TNO_DistDist}}
\end{figure}

For each entry in the catalog of SDSS images possibly
containing an observation of a
Solar-System object with a predicted magnitude
brighter than $V_{{\rm pred}}=22.0$ (\S\ref{Im}),
I downloaded the SDSS source catalog corresponding to
that image\footnote{These are the tsObj files stored in http://das.sdss.org/imaging/ which are described in http://www.sdss.org/dr7.1/dm/flatFiles/tsObj.html}.
Then
I searched for all the SDSS sources within $8''$ of the predicted position of the object.
Figure~\ref{fig:TNO_DistDist} shows the distribution of the angular separations
between the 
predicted object position and the nearest SDSS source.

In order to remove photometric measurements contaminated 
by nearby sources and possible
false detections, I selected only sources which have exactly
one SDSS match within $8''$ of the outer Solar-System object
predicted position.
I further demanded that this SDSS source is
within $1.5''$ from the predicted position of the object.
Moreover, I selected only sources for which there are
no USNO-B1 (Monet et al. 2003) objects within $8''$
from the position of the SDSS matched source.
Finally, I removed from the list of SDSS matches entries with
$r$-band magnitude errors larger than 0.2\,mag.
The final list contains 388 probable identifications of
42 unique outer Solar-System objects in SDSS images.
Each of these objects has between 1 and 49 measurements.
Table~\ref{Table_Ast_PerObsTNO} presents the astrometric properties of the 388 identifications,
while Table~\ref{Table_Phot_PerObsTNO} lists all the predicted and measured photometry.
\begin{deluxetable*}{llllllrrrrl}
\tablecolumns{11}
\tablewidth{0pt}
\tablecaption{Astrometric measurements of Solar-System objects identified in SDSS images}
\tablehead{
\colhead{Name} &
\colhead{JD-2450000} &
\colhead{Run} &
\colhead{Rerun} &
\colhead{Camcol} &
\colhead{Field} &
\colhead{$\alpha_{2000}^{obs}$} &
\colhead{$\delta_{2000}^{obs}$} &
\colhead{$\alpha_{2000}^{pred}$} &
\colhead{$\delta_{2000}^{pred}$} &
\colhead{Dist} \\
\colhead{} &
\colhead{day} &
\colhead{} &
\colhead{} &
\colhead{} &
\colhead{} &
\colhead{deg} &
\colhead{deg} &
\colhead{deg} &
\colhead{deg} &
\colhead{arcsec} 
}
\startdata
24835  &  1464.856934 &   1035&  40& 4&  126&   24.909451&  $14.267754$&  24.909542&  $14.267833$& 0.43\\
26375  &  2318.829706 &   2986&  40& 3&  273&  159.458594&  $ 6.417053$& 159.458667&  $ 6.417111$& 0.34\\
       &  2338.709340 &   3015&  40& 3&  300&  159.008464&  $ 6.578481$& 159.008458&  $ 6.578528$& 0.17\\
35671  &  1819.838123 &   1755&  40& 2&  340&  354.611093&  $-0.825181$& 354.611083&  $-0.825139$& 0.16\\
       &  2196.787392 &   2649&  40& 2&   94&  355.871559&  $-0.418224$& 355.871583&  $-0.418222$& 0.09
\enddata
\tablecomments{Astrometric measurements of 388 identifications of 42 outer Solar-System objects.
Name is the minor planet number or provisional designation, Run/Rerun/Camcol/Field identifies
the unique SDSS $ugriz$ image, while JD provides the time at which the $r$-band image was obtained.
$\alpha$ and $\delta$ are the J2000.0 coordinates of the object,
while superscript ``obs'' is for observed coordinates, and superscript ``pred' is for predicted coordinates.
Dist is the angular distance between the observed and predicted coordinates.
This table is published in its entirety in the electronic edition of
the {\it Astrophysical Journal}. A portion of the full table is shown here for
guidance regarding its form and content.
}
\label{Table_Ast_PerObsTNO}
\end{deluxetable*}
\begin{deluxetable*}{lllllllllllllllll}
\tablecolumns{17}
\tablewidth{0pt}
\tablecaption{Photometric properties of Solar-System objects identified in SDSS images}
\tablehead{
\colhead{Name} &
\colhead{JD-2450000} &
\colhead{$R$} &
\colhead{$\Delta$} &
\colhead{$\beta$} &
\colhead{$V_{{\rm pred}}$} &
\colhead{Type} &
\colhead{$u$} &
\colhead{$g$} &
\colhead{$r$} &
\colhead{$i$} &
\colhead{$z$} &
\colhead{$\Delta{u}$} &
\colhead{$\Delta{g}$} &
\colhead{$\Delta{r}$} &
\colhead{$\Delta{i}$} &
\colhead{$\Delta{z}$} \\
\colhead{}    &
\colhead{day} &
\colhead{AU}  &
\colhead{AU}  &
\colhead{deg} &
\colhead{mag} &
\colhead{}    &
\colhead{mag} &
\colhead{mag} &
\colhead{mag} &
\colhead{mag} &
\colhead{mag} &
\colhead{mag} &
\colhead{mag} &
\colhead{mag} &
\colhead{mag} &
\colhead{mag} 
}
\startdata
24835  & 1464.856934&  39.497&  38.512&  0.24& 20.8&  6 &\nodata& 20.879& 20.363& 20.192& 19.930&  0.271& 0.038& 0.034& 0.043& 0.145\\
26375  & 2318.829706&  34.179&  33.221&  0.42& 20.1&  6 &\nodata& 20.859& 20.087& 19.755& 19.634&  0.269& 0.033& 0.025& 0.026& 0.059\\
       & 2338.709340&  34.191&  33.206&  0.18& 20.0&  6 &\nodata& 20.929& 20.161& 19.784& 19.592&  0.605& 0.034& 0.027& 0.026& 0.064\\
35671  & 1819.838123&  38.156&  37.188&  0.38& 21.6&  6 &\nodata& 21.760& 21.262& 21.140&\nodata&  0.425& 0.076& 0.062& 0.074& 0.275\\
       & 2196.787392&  38.109&  37.203&  0.63& 21.7&  6 &\nodata& 21.653& 21.270& 21.291&\nodata&  1.087& 0.063& 0.057& 0.079& 0.242
\enddata
\tablecomments{As in Table~\ref{Table_Ast_PerObsTNO} except for the photometric properties.
$V_{{\rm pred}}$ is the minor planet predicted $V$-band magnitude at the time of the SDSS observation.
Type is the SDSS morphological classification (6: star; 3: galaxy).
$ugriz$ are the SDSS magnitudes, while their corresponding errors are
$\Delta{u}$, $\Delta{g}$, $\Delta{r}$, $\Delta{i}$ and $\Delta{z}$.
In cases in which the error magnitudes
are larger than 0.2\,mag,
I replaced the SDSS magnitude with the ``no-data'' symbol.
However, I kept the errors in the table.
Note that the absolute planetary $r$-band magnitude of measurement number 18 of object 145452,
measurement number 7 of object 145453,
and measurement number 13 of object 145480 deviate by more than one
magnitude from the median absolute planetary magnitude
and are probably bad measurements.
These measurements are listed in this table but are not used in
the subsequent analysis (e.g., they are not shown in Figure~\ref{fig:TNO_r_beta}
and they are excluded from the phase--angle slope parameter fits summarized in Table~\ref{Table_MeanSlope}).
This table is published in its entirety in the electronic edition of
the {\it Astrophysical Journal}. A portion of the full table is shown here for
guidance regarding its form and content.}
\label{Table_Phot_PerObsTNO}
\end{deluxetable*}

I also calculated the absolute planetary magnitude\footnote{Defined
as the magnitude of an object observed at opposition and
at 1\,AU from the Sun and Earth.}, neglecting phase effects (see \S\ref{Var}),
$H_{f}=m_{f}-5\log_{10}(R\Delta)$,
where $m_{f}$ is the apparent magnitude in band $f$
($u$, $g$, $r$, $i$ or $z$), $R$ is the object's
heliocentric
distance\footnote{Denoted by $R$ to distinguish it from the SDSS $r$-band magnitude.},
and $\Delta$ is its geocentric distance.
The values  of $R$, $\Delta$ and the
phase angle $\beta$
(defined as the Sun--target--observer angle), for each observation,
are listed in Table~\ref{Table_Phot_PerObsTNO}.
I also calculated the median, standard deviation (StD),
and range 
of the absolute magnitude measurements (Table~\ref{Table_Phot_PerObsTNO}).
\begin{deluxetable*}{lllllllllllllrrr}
\tablecolumns{16}
\tablewidth{0pt}
\tablecaption{Mean photometric properties of outer Solar-System objects identified in SDSS images}
\tablehead{
\colhead{Name} &
\colhead{$N_{obs}^{r}$} &
\colhead{$H_{u}$} &
\colhead{$H_{g}$} &
\colhead{$H_{r}$} &
\colhead{$H_{i}$} &
\colhead{$H_{z}$} &
\colhead{$u_{StD}$} &
\colhead{$g_{StD}$} &
\colhead{$r_{StD}$} &
\colhead{$i_{StD}$} &
\colhead{$z_{StD}$} &
\colhead{$r_{range}$} &
\colhead{$a$} &
\colhead{$e$} &
\colhead{$I$} \\
\colhead{} &
\colhead{} &
\colhead{mag} &
\colhead{mag} &
\colhead{mag} &
\colhead{mag} &
\colhead{mag} &
\colhead{mag} &
\colhead{mag} &
\colhead{mag} &
\colhead{mag} &
\colhead{mag} &
\colhead{mag} &
\colhead{AU} &
\colhead{} &
\colhead{deg} 
}
\startdata
24835               &   1& \nodata&  4.968&  4.452&  4.281&  4.019&\nodata&   0.000 &   0.000 &   0.000 &   0.000  &   0.000 &    41.957& 0.106& 27.000\\
26375               &   2& \nodata&  5.619&  4.848&  4.494&  4.337&\nodata&   0.050 &   0.052 &   0.021 &   0.029  &   0.073 &    55.108& 0.415&  7.631\\
35671               &  16& \nodata&  6.079&  5.573&  5.450&  4.975&\nodata&   0.104 &   0.106 &   0.099 &   0.028  &   0.398 &    38.110& 0.043&  4.596\\
38628 (Huya)        &   1& \nodata&  5.600&  4.775&  4.394&  4.249&\nodata&   0.000 &   0.000 &   0.000 &   0.000  &   0.000 &    39.373& 0.276& 15.488\\
65489 (Ceto)        &   1& \nodata&  7.182&  6.519&  6.136&  5.893&\nodata&   0.000 &   0.000 &   0.000 &   0.000  &   0.000 &    99.676& 0.821& 22.325\\
73480               &  11& \nodata&  9.270&  8.641&  8.337&  8.151&\nodata&   0.105 &   0.074 &   0.079 &   0.114  &   0.238 &    31.245& 0.572& 16.627\\
79360               &   4& \nodata&  6.187&  5.287&  4.773&\nodata&\nodata&   0.132 &   0.155 &   0.054 & \nodata  &   0.325 &    43.739& 0.008&  2.250\\
82075               &   3& \nodata&  5.222&  4.496&  4.215&  4.084&\nodata&   0.153 &   0.119 &   0.060 &   0.199  &   0.225 &    57.686& 0.288& 19.840\\
82155               &   1& \nodata&  6.634&  6.141&  5.693&\nodata&\nodata&   0.000 &   0.000 &   0.000 & \nodata  &   0.000 &    84.625& 0.617& 12.739\\
82158               &   1& \nodata&  6.963&  6.129&  5.708&\nodata&\nodata&   0.000 &   0.000 &   0.000 & \nodata  &   0.000 &   212.866& 0.839& 30.842\\
90482 (Orcus)       &   2&   4.110&  2.702&  2.222&  2.082&  2.054&  0.021&   0.050 &   0.070 &   0.018 &   0.075  &   0.098 &    39.173& 0.227& 20.573\\
119878              &   2& \nodata&  7.120&  6.216&  5.674&  5.198&\nodata&   0.135 &   0.040 &   0.046 &   0.000  &   0.056 &    53.648& 0.344& 15.759\\
120132              &   1& \nodata&  5.436&  4.653&  4.265&  4.331&\nodata&   0.000 &   0.000 &   0.000 &   0.000  &   0.000 &    49.197& 0.247& 11.798\\
120181              &   2& \nodata&  8.118&  7.309&  6.845&\nodata&\nodata&   0.043 &   0.075 &   0.043 & \nodata  &   0.106 &    32.536& 0.177&  2.717\\
135182              &   1& \nodata&\nodata&  6.965&\nodata&\nodata&\nodata& \nodata &   0.000 & \nodata & \nodata  &   0.000 &    37.285& 0.021&  1.837\\
139775              &  10& \nodata&  7.806&  7.047&  6.593&\nodata&\nodata&   0.067 &   0.197 &   0.225 & \nodata  &   0.644 &    39.649& 0.199&  6.480\\
144897              &  17& \nodata&  5.148&  4.315&  4.026&  3.830&\nodata&   0.086 &   0.069 &   0.232 &   0.172  &   0.223 &    39.203& 0.043&  9.524\\
145451              &  27&   5.996&  4.852&  4.436&  4.289&  4.283&  0.251&   0.084 &   0.070 &   0.084 &   0.126  &   0.321 &    91.720& 0.617& 28.759\\
145452              &  49&   5.922&  4.489&  3.694&  3.370&  3.206&  0.000&   0.092 &   0.075 &   0.067 &   0.090  &   0.375 &    41.759& 0.028& 19.236\\
145453              &  36&   5.740&  4.454&  4.003&  3.898&  3.843&  0.331&   0.067 &   0.037 &   0.053 &   0.148  &   0.141 &    43.422& 0.143& 28.509\\
145480              &  40& \nodata&  5.235&  4.451&  4.119&  3.722&\nodata&   0.143 &   0.133 &   0.100 &   0.154  &   0.852 &    76.591& 0.397& 26.429\\
150642              &   2& \nodata&  6.463&  5.935&  5.738&\nodata&\nodata&   0.036 &   0.143 &   0.409 & \nodata  &   0.202 &    45.019& 0.116& 10.235\\
229762              &   1& \nodata&  4.221&  3.451&  3.093&  2.936&\nodata&   0.000 &   0.000 &   0.000 &   0.000  &   0.000 &    73.744& 0.490& 23.367\\
2000 CN105          &   1& \nodata&\nodata&  5.498&  5.252&\nodata&\nodata& \nodata &   0.000 &   0.000 & \nodata  &   0.000 &    44.361& 0.100&  3.422\\
2002 KY14           &   2& \nodata& 11.362& 10.678&  9.922&  9.651&\nodata&   0.043 &   0.299 &   0.089 &   0.127  &   0.423 &    12.632& 0.318& 19.452\\
2002 PQ152          &   1& \nodata&\nodata&  8.886&  8.595&\nodata&\nodata& \nodata &   0.000 &   0.000 & \nodata  &   0.000 &    25.930& 0.192&  9.334\\
2002 QX47           &  12& \nodata&  9.341&  8.851&  8.587&  8.096&\nodata&   0.176 &   0.128 &   0.103 &   0.000  &   0.383 &    25.604& 0.375&  7.264\\
2003 QW90           &  11& \nodata&  5.914&  5.069&  4.588&  4.275&\nodata&   0.124 &   0.118 &   0.096 &   0.119  &   0.378 &    44.024& 0.075& 10.337\\
2003 UZ413          &   6&   5.723&  4.827&  4.259&  4.044&  4.005&  0.000&   0.106 &   0.042 &   0.037 &   0.081  &   0.118 &    39.401& 0.223& 12.044\\
2004 PG115          &  13& \nodata&  5.903&  5.080&  4.686&  4.459&\nodata&   0.085 &   0.151 &   0.057 &   0.151  &   0.568 &    91.908& 0.604& 16.277\\
2004 VT75           &   1& \nodata&\nodata&  6.249&  5.644&  4.961&\nodata& \nodata &   0.000 &   0.000 &   0.000  &   0.000 &    39.544& 0.213& 12.818\\
2005 CB79           &   1& \nodata&  5.149&  4.680&  4.558&  4.487&\nodata&   0.000 &   0.000 &   0.000 &   0.000  &   0.000 &    43.167& 0.139& 28.664\\
2005 RO43           &  32& \nodata&  7.821&  7.193&  6.956&  6.500&\nodata&   0.138 &   0.097 &   0.149 &   0.000  &   0.429 &    28.880& 0.518& 35.415\\
2005 RS43           &  47& \nodata&  5.691&  5.030&  4.781&  4.448&\nodata&   0.129 &   0.114 &   0.102 &   0.179  &   0.579 &    48.154& 0.203&  9.995\\
2006 QP180          &   5& \nodata& 10.385&  9.827&  9.215&  8.726&\nodata&   0.216 &   0.244 &   0.229 &   0.002  &   0.597 &    38.600& 0.658&  4.953\\
2006 SX368          &   7& \nodata& 10.338&  9.694&  9.430&  9.081&\nodata&   0.117 &   0.054 &   0.092 &   0.063  &   0.146 &    22.293& 0.463& 36.283\\
2007 RT15           &   4& \nodata&  7.213&  6.537&  6.274&\nodata&\nodata&   0.153 &   0.166 &   0.184 & \nodata  &   0.361 &    39.662& 0.234& 12.924\\
2007 TG422          &   5& \nodata&  7.079&  6.305&  6.194&\nodata&\nodata&   0.042 &   0.193 &   0.061 & \nodata  &   0.493 &   549.606& 0.935& 18.601\\
2007 TK422          &   2& \nodata&  9.824&  9.061&  9.128&\nodata&\nodata&   0.000 &   0.047 &   0.000 & \nodata  &   0.067 &    21.264& 0.198&  3.066\\
2007 UM126          &   2& \nodata& 11.121& 10.700& 10.559& 10.067&\nodata&   0.141 &   0.504 &   0.331 &   0.000  &   0.713 &    12.919& 0.340& 41.698\\
2007 VH305          &   4& \nodata& 12.511& 12.296& 11.961&\nodata&\nodata&   0.245 &   0.151 &   0.226 & \nodata  &   0.328 &    24.553& 0.666&  6.205\\
2008 QB43           &   1& \nodata&\nodata&  5.065&  4.603&  3.874&\nodata& \nodata &   0.000 &   0.000 &   0.000  &   0.000 &    43.401& 0.220& 26.354
\enddata
\tablecomments{Columns description: $N_{obs}^{r}$ is the number of $r$-band observations in Table~\ref{Table_Phot_PerObsTNO};
$H_{u}$ through $H_{z}$ are the median absolute planetary magnitudes,
not corrected for phase angle, in the $ugriz$ bands;
$u_{std}$ through $z_{std}$ are the StD in $H_{u}$ through $H_{z}$, respectively,
but after removing the three bad measurements (see Table~\ref{Table_Phot_PerObsTNO});
and $r_{range}$ is the range in $H_{r}$ over all measurements, excluding the three bad measurements.}
\label{Table_MeanPhotProp}
\end{deluxetable*}

\subsection{Verification}
\label{Varif}

As shown in Figure~\ref{fig:TNO_DistDist},
a large fraction of the
matched SDSS sources are found
within $1.5''$ from the predicted position of the
Solar-System objects.
In contrast, the probability for a match with background sources
should increase as the square of the distance.
Since the number of matches above $1.5''$ is small
I argue that the fraction of false identification
and contaminated photometry in Tables~\ref{Table_Ast_PerObsTNO}--\ref{Table_Phot_PerObsTNO}
is negligible.
Furthermore, I note that assuming a source density of $10^{4}$\,deg$^{-2}$ in the SDSS images,
the probability of finding a source within $1.5''$ from a random position is $0.5\%$.

Nevertheless, as an additional test I uploaded cutouts
of the SDSS images containing
some of the candidates, along with SDSS images of the same sky
positions taken at different epochs. If such an extra epoch
image was not available then I uploaded instead an image from the
Palomar Sky
Survey\footnote{http://archive.stsci.edu/cgi-bin/dss\_form} (Reid et al. 1991).
I inspected by eye about 100 of these cutouts and verified
that the Solar-System object candidate is indeed detected only in one epoch.
Finally, I note that the differences between the predicted and measured magnitudes
(Table~\ref{Table_Phot_PerObsTNO}) is typically small, on the order of $0.3$\,mag.

In Table~\ref{Table_Phot_PerObsTNO} I list also the SDSS morphological type
(6: star; 3: galaxy) for the TNOs.
Some of the sources are identified as possible resolved objects.
This is presumably because the reliability
of SDSS star--galaxy separation degrades near
the survey detection limit.

\section{Object color}
\label{Color}

\begin{figure}
\centerline{\includegraphics[width=8.5cm]{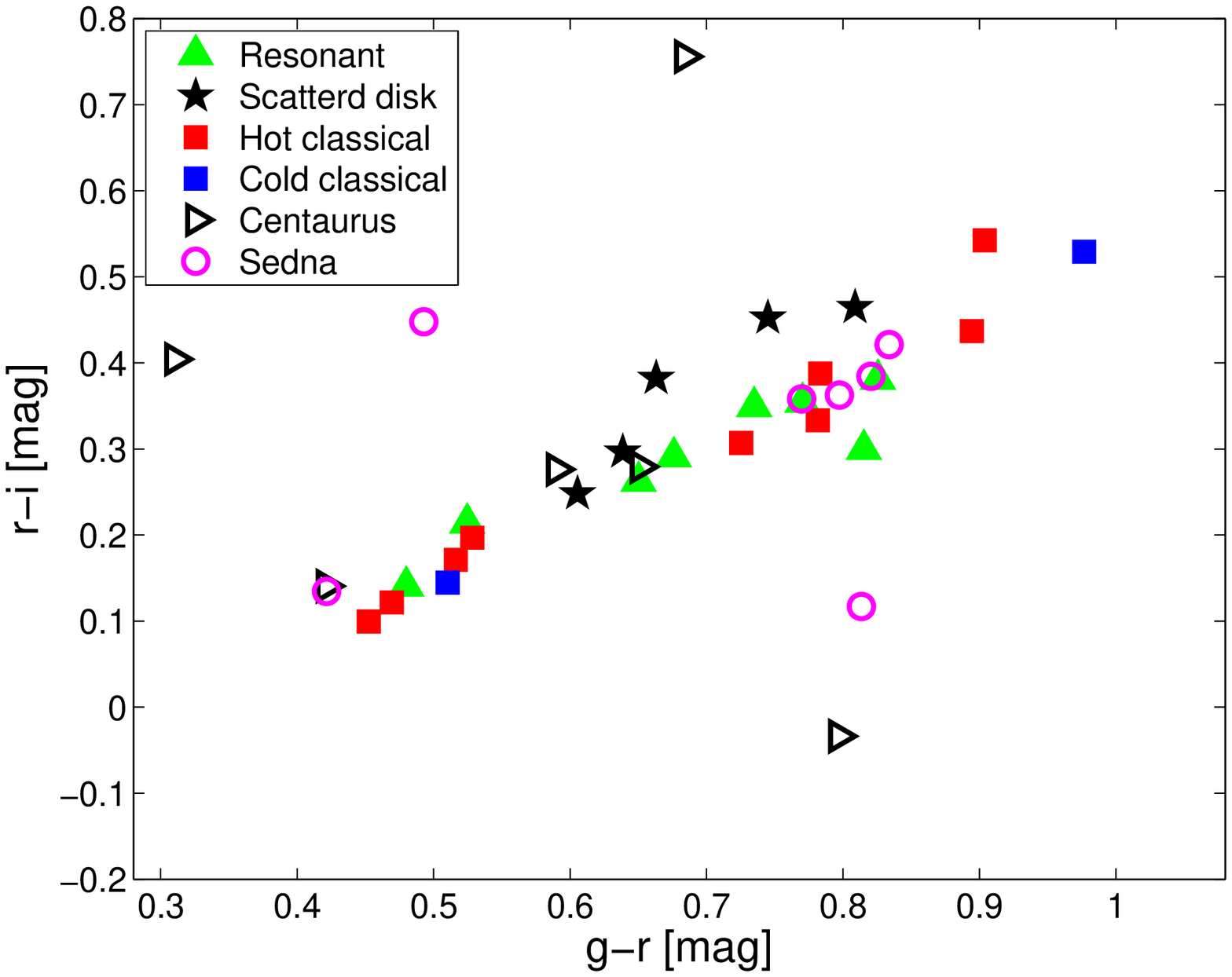}}
\caption{The $g-r$ vs. $r-i$ color--color diagram for 37 objects.
Different symbols represent distinct classes of objects.
Here, I define resonant objects as those having orbital periods
within 2.5\% of $3/2$, $4/3$, $2$, or $5/2$ times the Neptune orbital period.
Scattered disk objects are defined as those having $a>28$\,AU, $q<32$\,AU, $Q>32$\,AU
and that are not resonant objects.
Cold classical objects are defined  as those having $a>30$\,AU, $a<60$\,AU, $I<6$\,deg
and not being resonant or scattered disk objects.
Hot classical are defined similarly to cold classical
but having $I>6$\,deg.
Centaurus objects are defined as those having $28\,{\rm AU}>a>5.204$\,AU,
and ``Sedna'-like objects are all the objects with $a>60$\,AU
that are neither resonant objects nor scattered disk objects.
I note that some of the various subclasses are not well defined.
\label{fig:Hgr_Hri}}
\end{figure}

The absolute planetary magnitude of objects identified in SDSS images
are listed in Table~\ref{Table_MeanPhotProp}.
If multiple-epoch observations are available,
I adopt the median of the object magnitudes over all epochs
as the object's magnitude.
In cases in which an object was observed in multiple epochs,
I also give the standard deviation (StD) of the absolute magnitudes,
and the range of the $r$-band absolute magnitudes.
I note that the variability indicators in this table do~not separate
between variability due to phase-angle variations and other causes (e.g., rotation).
Separation of phase-angle and rotation-induced variability is possible
only when a large number of observations is available.
Nevertheless, the variability indicators in this table
give a rough idea regarding which objects may be variable
and which objects are less likely to be variable.
For objects which have more than ten observations more reliable variability
indicators, which are calculated after subtracting the phase-angle variations,
are available in Table~5.
Figure~\ref{fig:Hgr_Hri} shows the $g-r$ vs. $r-i$ color--color diagram for the 37 objects
for which both $g-r$ and $r-i$ color measurements are available.
The symbols indicate different subclasses of objects
(see figure legend and caption).
I note that the $g-r$ vs. $r-i$ locus of objects in Figure~\ref{fig:Hgr_Hri}
is generally similar, but offset, relative to the $B-V$ vs. $R-I$ colors locus of TNOs
(for comparison see Fig.~2 in Tegler \& Romanishin 2003).

In order to explore possible correlations of the colors
with orbital properties I calculated
the Spearman rank correlation coefficients between
the $g-r$, $r-i$, and $g-i$ colors and various
photometric and orbital properties of these objects
and their subgroups
(e.g., the groups listed in Fig.~\ref{fig:Hgr_Hri}).
This approach has the disadvantage that it introduce many
trials, and reduces the significance of any reported correlation
by a complicated way that depends on the correlations
between the various parameters.
Nevertheless, this may give us some ideas about
where to look for correlations when larger samples,
based on the same filters, become available.
For each correlation I also calculated the
probability of getting a value larger than the correlation
coefficient.
This was calculated
from the
correlation coefficients'
probability distributions derived from $10^{4}$ bootstrap simulations
(Efron 1982; Efron \& Tibshirani 1993).
In each simulation, considering two lists ``$X$'' and ``$Y$'',
I select for each entry in $X$
a random entry in $Y$
and calculate the correlation between the two randomly permuted lists.
I calculated the correlations of the three colors
$g-r$, $r-i$, and $g-i$
with
the $r-i$ color,
$H_{r}$,
semi-major axis $a$,
orbital eccentricity $e$,
orbital inclination $I$,
perihelion distance $q$,
aphelion distance $Q$,
the orbital angular momentum in the
$z$-direction ($L_{z}=\sqrt{a (1-e^{2})}\cos{I}$),
and the Tisserand parameter calculated with respect
to Neptune ($T_{N}=\frac{a_{N}}{a}+2\sqrt{\frac{a}{a_{N}}(1-e^{2})}\cos{I}$,
where $a_{N}=30.104$\,AU is the orbital semi-major axis of Neptune).

The only significant correlation reported in the literature
is the color-inclination correlation
(e.g., Trujillo \& Brown; Peixinheo et al. 2008).
Here, I find only weak evidence for this correlation.
Specifically, I find that the $g-r$ colors
have a correlation coefficient of $-0.90$ and $-0.68$
for scattered disk objects and hot KBOs, respectively.
The probability of getting correlation coefficient
which are smaller than these values are $1.4\%$
and $2.3\%$ (per trial), respectively.
For the rest of the populations investigated here
the $g-r$-inclination correlation one-sided false alarm probability
is larger than $2.5\%$ (corresponds to $2\sigma$).
The correlation I find is weaker and less significant than
that found in other studies.
Possible explanations for the differences between the 
correlations found in this paper and in other works are:
(i) the different filters used by different studies;
(ii) selection biases that plague the various samples;
and (iii) the small sample size.

The nature of the color--inclination correlation
is not clear.
Among the possible explanations are collisional resurfacing (Luu \& Jewitt 1996;
Jewitt \& Luu 2001) in which
collisions between TNOs expose
fresh material and change their colors
and at the same time excite their inclinations.
Another possibility is that the colors of KBOs are primordial
and related to dynamical groupings.
However, both explanation have been
challenged by observations
(see Trujillo \& Brown 2002; Volk \& Malhotra 2011).

Peixinho et al. (2008) found that there is a break
in the ``relation'' between color and orbital inclination,
where objects with $I\ltorder 12$\,deg shows no correlation
with color.
Moreover, the perihelion distance ($=a[1-e]$)
and inclination of Classical KBOs are known to be loosely related.
I note that the correlation of color with
inclination and perihelion distance
have a functional
resemblance to the functional form of the orbital angular momentum
in the $z$-direction
($L_{z}$)\footnote{$L_{z}$ depends on $\cos(I)$ which varies
by only $2$\% between 0 and 12\,deg.}.
Curiously, I find that the correlation
between the $g-r$ color and $L_{z}$ has a one-sided false
alarm probability of $0.4\%$.
It will be interesting to test this correlation
using larger samples.

I note that collisions 
between objects conserve the total angular momentum
of the bodies involve in the collision.
Since the angular momentum
of individual bodies is not conserved,
the absolute value of the angular momentum of individual bodies
{\it may} (at least for mostly elastic collisions) statistically increase after a collision.
However, this depends on the details of the collisions
(e.g., elasticity and initial orbits).
Therefore,
I cannot rule out that the $L_{{\rm z}}$--color correlation, if real,
is a byproduct of TNO collisions.
At this stage, it is not clear that the color--$L_{z}$
correlation really has a physical meaning rather
than being a combination of several (physically unrelated) correlations
(e.g., color--inclination, and perihelion distance--inclination correlations).

\section{Variability}
\label{Var}

Thirteen objects in my sample have ten or more SDSS $r$-band measurements
with errors smaller than 0.2\,mag.
Although the observations are too sparse to unambiguously identify
periods, they are good enough  to study the objects'
reflectivity as a function of phase angle (\S\ref{PhaseAngle})
and to search for large amplitude variability due to rotation and binarity (\S\ref{Rot}).

\subsection{Phase angle variations}
\label{PhaseAngle}

Solar-System bodies are known to vary in brightness with
phase angle.
There are two important physical reasons for this variation.
The first is shadow hiding, in which particles on the planetary surface
cast shadows on adjacent areas: the shaded area is minimized near opposition.
The second is an interference mechanism called coherent backscatter in which
reflected light, depending on the regolith properties,
may constructively interfere, resulting in an increased brightness
at opposition (Hapke 1993; 2002).

\begin{figure*}
\centerline{\includegraphics[width=16cm]{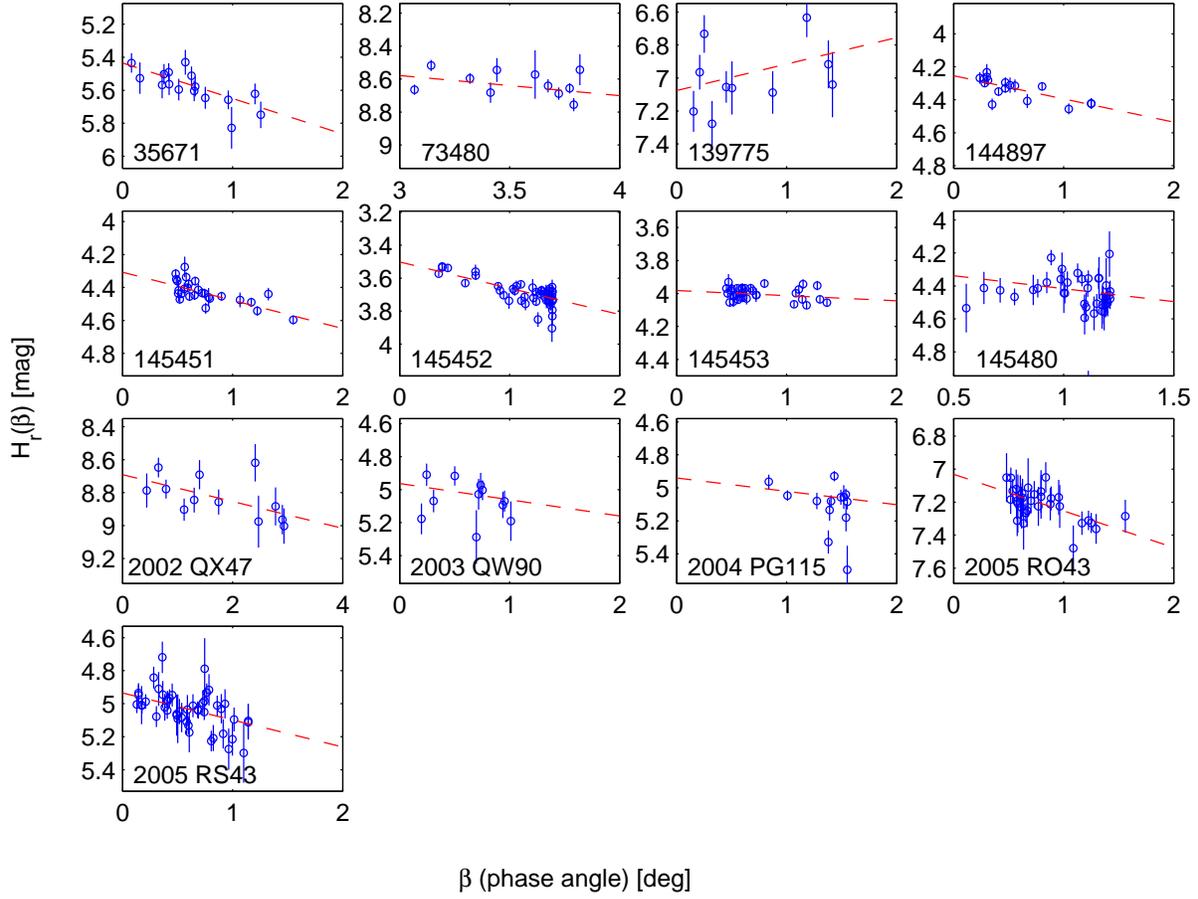}}
\caption{Absolute planetary $r$-band magnitude as a function of
phase angle ($\beta$). The negative slope parameter of 139775
is presumably due to variability.
\label{fig:TNO_r_beta}}
\end{figure*}

Hapke (2002) presented models of these
effects. These models have seven degrees of freedom.
Given 
the relatively small number of observations and limited range
of phase angles in which the SDSS observations
were obtained, I fit a linear relation of the form
\begin{equation}
H_{f}(\beta)= H_{f,0} + S_{f}\beta.
\label{LinSlope}
\end{equation}
Here, $f$ is the filter name ($g$, $r$ or $i$),
$H_{f,0}$ is the absolute planetary magnitude at zero phase angle,
$S_{f}$ is a linearized phase angle slope parameter for filter $f$,
and $\beta$ is the phase angle.
Figure~\ref{fig:TNO_r_beta} shows $H_{r}(\beta)$ as a function of $\beta$ for these 13 objects.
As seen in other Solar-System objects,
these 13 objects with one exception, are brightest near opposition.
To quantify this, in Table~\ref{Table_MeanSlope}
I summarize the phase angle slope parameters and related
information for the 13 objects.
The only object that does not follow this rule
is 139775.
A plausible explanation is that this
object has a large intrinsic variability due to
rotation or binarity (see \S\ref{Rot}).
I note that the fits are performed only for measurements
which $H_{r}(\beta)$ is within 1\,magnitude
of the median of $H_{r}(\beta)$.
The three measurements which
do~not fulfill this condition,
and were removed, are indicated
in Table~\ref{Table_Phot_PerObsTNO}.
\begin{deluxetable*}{lrrllrrll}
\tablecolumns{9}
\tablewidth{0pt}
\tablecaption{Linearized phase angle slope parameters for objects with multiple observations}
\tablehead{
\colhead{Name} &
\colhead{$H_{r,0}$} &
\colhead{$S_{r}$} &
\colhead{$\chi^{2}_{r}$} &
\colhead{$dof_{r}$} &
\colhead{$H_{g,0}$} &
\colhead{$S_{g}$} &
\colhead{$\chi^{2}_{g}$} &
\colhead{$dof_{g}$} \\
\colhead{} &
\colhead{mag} &
\colhead{mag\,deg$^{-1}$} &
\colhead{} &
\colhead{} &
\colhead{mag} &
\colhead{mag\,deg$^{-1}$} &
\colhead{} &
\colhead{} 
}
\startdata
35671      &$ 5.43\pm 0.04$&$ 0.215\pm 0.053$& 10.6& 14 &$ 6.02\pm 0.04$&$ 0.111\pm 0.058$& 27.5& 14\\ 
73480      &$ 8.21\pm 0.15$&$ 0.122\pm 0.043$& 21.6&  9 &$ 8.83\pm 0.16$&$ 0.128\pm 0.046$& 28.4&  8\\ 
139775     &$ 7.08\pm 0.07$&$-0.161\pm 0.092$& 20.0&  8 &$ 7.66\pm 0.30$&$ 0.224\pm 0.484$&  0.0&  0\\ 
144897     &$ 4.25\pm 0.01$&$ 0.141\pm 0.023$& 28.4& 15 &$ 5.03\pm 0.02$&$ 0.195\pm 0.030$& 29.4& 15\\ 
145451     &$ 4.31\pm 0.02$&$ 0.171\pm 0.020$& 61.9& 25 &$ 4.78\pm 0.02$&$ 0.096\pm 0.022$&121.6& 25\\ 
145452     &$ 3.50\pm 0.01$&$ 0.160\pm 0.011$& 91.1& 46 &$ 4.26\pm 0.02$&$ 0.184\pm 0.017$&108.2& 45\\ 
145453     &$ 3.98\pm 0.01$&$ 0.031\pm 0.015$& 65.6& 33 &$ 4.49\pm 0.01$&$-0.038\pm 0.017$&125.1& 33\\ 
145480     &$ 4.26\pm 0.08$&$ 0.157\pm 0.078$& 70.2& 37 &$ 5.24\pm 0.12$&$-0.035\pm 0.107$& 59.8& 32\\ 
2002 QX47  &$ 8.69\pm 0.05$&$ 0.081\pm 0.029$& 15.1& 10 &$ 9.20\pm 0.06$&$ 0.116\pm 0.036$&  9.1&  9\\ 
2003 QW90  &$ 4.96\pm 0.06$&$ 0.098\pm 0.085$& 14.4&  9 &$ 5.92\pm 0.09$&$-0.028\pm 0.142$&  5.9&  6\\ 
2004 PG115 &$ 4.94\pm 0.07$&$ 0.081\pm 0.054$& 48.4& 11 &$ 6.05\pm 0.13$&$-0.112\pm 0.096$&  7.0& 10\\ 
2005 RO43  &$ 7.03\pm 0.05$&$ 0.222\pm 0.056$& 18.9& 30 &$ 7.66\pm 0.06$&$ 0.155\pm 0.067$& 40.5& 24\\ 
2005 RS43  &$ 4.94\pm 0.02$&$ 0.163\pm 0.038$& 57.6& 45 &$ 5.60\pm 0.03$&$ 0.126\pm 0.047$& 76.3& 38
\enddata
\tablecomments{Mean photometric properties and slope parameters
for the 13 outer Solar-System objects with more than nine SDSS
observations.
The fits are performed only using measurements with
photometric errors smaller than 0.2\,mag.
Therefore, the $g$-band and $r$-band slope measurements
are not always based on data points taken at the same epochs.
Column descriptions:
$H_{f,0}$ is the best fit absolute planetary magnitude at zero phase
angle for filter $f$;
$S_{f}$ is the linearized slope parameter;
$\chi^{2}_{f}$ and $dof_{f}$ indicate
the $\chi^{2}$ and the number of degrees of freedom of the best fit for filter $f$.}
\label{Table_MeanSlope}
\end{deluxetable*}
Interestingly, in most cases the slope parameter $H_{r}(\beta)$ is larger than
0.04\,mag\,deg$^{-1}$.
Such large slope parameters were argued to be the result
of coherent backscatter (see Schaefer, Rabinowitz \& Tourtellotte 2009).

\begin{figure}
\centerline{\includegraphics[width=8.5cm]{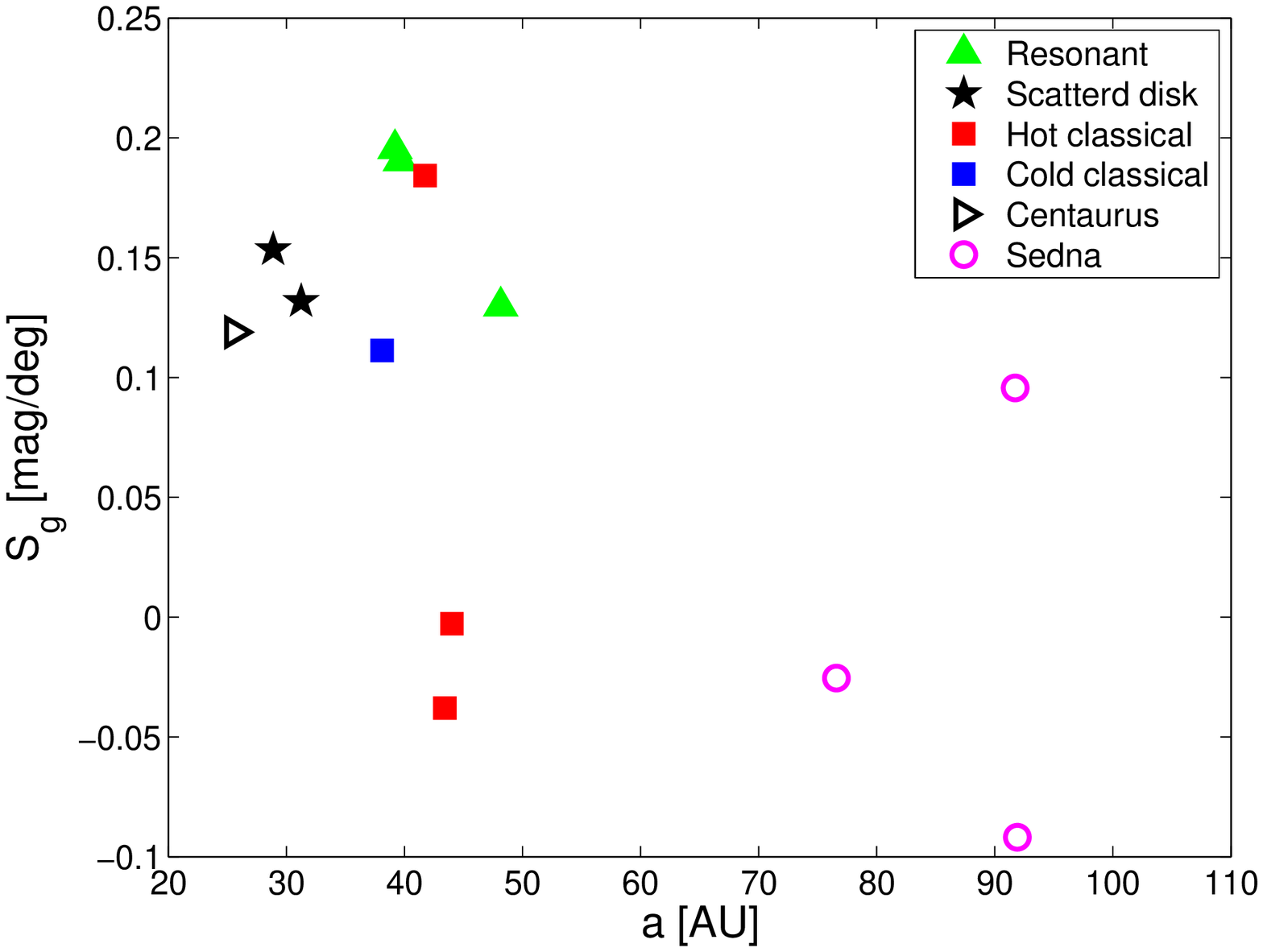}}
\caption{The linearized $g$-band slope parameter $S_{g}$
as a function of the semi-major axis $a$ of the 13 objects
with more than nine photometric measurements.
Symbols as in Fig.~\ref{fig:Hgr_Hri}.
The typical errors in $S_{g}$ are roughly $0.05$\,mag\,deg$^{-1}$ (see Table~\ref{Table_MeanSlope}). 
\label{fig:Slope_g_a}}
\end{figure}

Although the sample of objects for which I measure the slope parameters
is small I attempted to look for correlations between the slope parameters
and the orbital parameters of these TNOs.
Any hints for correlations found here can be tested in
the future using larger samples.
The only notable anti-correlations I find
are between $S_{g}$ and $a$, $Q$, and $P$,
where $P$ is the orbital period.
For example, the correlation between $S_{g}$ and $a$ is $-0.59$,
and the probability to get a correlation smaller than this
is 1.8\% per trial (roughly $2\sigma$ significance).
In order to test if this correlation is real,
larger samples are required.
Figure~\ref{fig:Slope_g_a} presents $S_{g}$ as a function
of $a$. This figure suggests that 
most of the apparent correlation arise due to a difference between
objects with large aphelion distances (i.e., ``Sedna''-like orbits)
and the rest of the population.

This anti-correlation means that the $g$-band slope parameter is steeper
for objects which are closer to the Sun.
However, this finding is based on a small sample of only 13 objects.
A possible selection bias that may cause such a correlation
is that closer objects are visible also on
larger phase angles, whereas further objects
are visible only near $\beta\approx0$.
Since in reality the derivative of the absolute-magnitude
phase-angle relation increases (in absolute value) near
opposition,
this may introduce the
observed anti-correlation between the slope
parameter and the semi-major axis.
However, most of the SDSS observations were taken
near a phase angle of $\beta\cong1$\,deg (see Figure~\ref{fig:TNO_r_beta}).
I also note that this selection bias should mostly
induce a correlation with $q$ rather than with $a$ or $Q$,
since objects with smaller $q$ are easier to detect.
Another possible caveat is that 
for three objects I find negative $g$-band slope parameters (see Table~\ref{Table_MeanSlope})
presumably affected by measurement errors and/or variability.
Therefore, more observations are required in order to confirm
the existence of such a correlation.

If this correlation is real, then there are several possible explanations:
(i) the regolith or surface composition properties
of TNOs vary with distance from the Sun;
or (ii) ``Sedna''-like objects have distinctive surface properties
which are related to their origin.
The variation in surface properties as a function of distance from the Sun
can originate, for example, 
if there are variations in the impact rate with micro-meteoroids
as a function of heliocentric distance,
or due to the crystallization properties of some ices.
Based on the Voyager-I and II spacecraft 
measurements, Gurnett et al. (2005) argued that the
number density of dust particles, as a function of heliocentric distance,
is roughly uniform
(up to distance of about 100\,AU).
Since the typical orbital speed of objects at 90\,AU
is 0.6 of that of objects at 30\,AU from the Sun,
this implies that the micro-meteoroids impact rate
does~not change dramatically for objects in my sample.

As different ices freeze at different temperatures,
the surface properties may also be affected by the
equilibrium temperature\footnote{The equilibrium temperature
of a reflective body in the Solar System is $\cong 278 (1-A)^{1/4} (R/1\,{\rm AU})^{-1/2}$\,K,
where $A$ is the object's geometric albedo and $R$ is its distance from the Sun.}
and escape velocity from the object (e.g., Schaller \& Brown 2007).
For an albedo of $A=0.04$, the equilibrium temperature varies between about 50\,K at 30\,AU
from the Sun to 29\,K at a heliocentric distance of about 90\,AU.
However, the three ``Sedna''-like objects shown in the right-hand side of
Figure~\ref{fig:Slope_g_a} are 145451, 145480 and 2004 PG115.
During the SDSS observations, these objects were near perihelion
at distances of 35, 46 and 36\,AU from the Sun, respectively.
Therefore, their actual equilibrium temperature,
at the time of observations, were similar to
those of some of the other objects in Figure~\ref{fig:Slope_g_a}.

I conclude that differences in surface properties
induced by the current orbit of these objects
are unlikely.
However,
I cannot rule out that ``Sedna''-like objects have a different origin
than some of the other classes of TNOs.

I note that Schaefer et al. (2009) found a significant
correlation between the slope parameter (near phase angle of $\beta\cong1$\,deg)
and the $B-I_{{\rm c}}$ color index and also a possible excluded region
in the slope parameter vs. inclination phase space.
I do~not find any indication for such correlations.
However, the Schaefer et al. (2009) sample is larger
(35 objects) and contains more diverse planetary objects
including the largest KBOs.

\subsection{Variability due to rotation and binarity}
\label{Rot}

Table~\ref{Table_MeanPhotProp} and Figure~\ref{fig:TNO_r_beta} suggest that some of the objects
in the catalog presented here are variable.
The variability of these sources may be the result of one
or more of the following reasons:
Small minor planets, probably with radii smaller than $\sim100$\,km,
may have irregular shapes
and therefore are variable.
Moreover, TNOs may show variations due to inhomogeneous surfaces.
Alternatively, fast rotation
of objects held by their own gravity
(i.e., radii larger than $\sim100$\,km)
may induce a highly non-spherical
equilibrium configuration and therefore
large amplitude variations (e.g., Leone et al. 1984; Rabinowitz et al. 2006).
Finally, contact binaries may show prominent eclipses
(e.g., Sheppard \& Jewitt 2004; Gnat \& Sari 2010).

Objects of 100\,km radius with an albedo of 0.04 (0.1)
will have absolute planetary
magnitude
of 7.6 (6.6).
Therefore, most
of the objects in Table~\ref{Table_MeanSlope} are probably larger than 100\,km.
In this case, it is probable that large amplitude variations
are either due to fast rotation, inhomogeneous surface albedo,
or binarity.
Since all these possibilities are interesting,
photometric follow-up observations of the 
most highly variable sources
in Tables~\ref{Table_MeanPhotProp} and \ref{Table_MeanSlope}
are desirable.

\section{Summary}
\label{Sum}

I cross-correlate SDSS observations with the ephemerides
of Solar-System bodies with $a>10$\,AU.
I present a catalog of SDSS photometric and astrometric
measurements of such minor planets based on SDSS observations.
After removing possible contaminated measurements.
I am left with 388 observations of 42 unique objects.

I find weak evidence for the previously reported
inclination-color correlation in
the scattered disk objects and hot classical KBOs.
I find marginally stronger correlation between the $g-r$ color
and orbital angular momentum in the $z$ direction, $L_{z}$, of the
entire population studied here.
I note that a correlation with $L_{z}$ is consistent
with the finding of Peixinho et al. (2008) that
objects with inclination below about 12\,deg shows
no color-inclination correlation.

Finally, the method presented here to collect photometric
observations of minor planets in surveys
which were not designed for Solar-System observations
can be utilized in other ongoing and planned surveys
such as the Palomar Transient Factory (Law et al. 2009; Rau et al. 2009),
Pan-STARRS (Kaiser et al. 2002),
SkyMapper (Keller et al. 2007),
and LSST (Tyson et al. 2003).

\acknowledgments
I thank Orly Gnat, Peter Goldreich,
Re'em Sari, and Hilke Schlichting 
for valuable discussions,
and I am greatfull to an anonymous referee for useful
suggestions.
EOO is supported by an Einstein fellowship
and NASA grants.

\end{document}